\documentstyle[11pt,IAU207_pasp,twoside,epsf]{article}
\def\edcomment#1{\iffalse\marginpar{\raggedright\sl#1\/}\else\relax\fi}
\reversemarginpar

\begin{document}
\title{Open, Massive and Globular Clusters -- Part of the same family?}
 \author{S{\o}ren S. Larsen}
\affil{Lick Observatory, University of California at Santa Cruz,
1156 High Street, Santa Cruz, CA 95064, USA}

\begin{abstract}
  Populations of young star clusters show significant differences even 
among ``normal'' disk galaxies.  In this contribution I discuss how 
properties of young cluster systems are related to those of their host 
galaxies, based on a recent study of clusters in a sample of 22 nearby 
spiral galaxies.  Luminous young clusters similar to the 
``super'' star clusters observed in starbursts and mergers exist in several 
of these galaxies, and it is found that the luminosity of the brightest star 
cluster as well as the specific luminosity of the cluster systems both 
correlate well with the host galaxy star formation rate.  When considering 
star clusters 
in different environments the traditional distinction between ``open'', 
``massive'' and ``globular'' clusters breaks down, underscoring the need for 
a universal physical description of cluster formation.  
\end{abstract}

\section{Introduction}

  It is becoming increasingly clear that ``open'' clusters as we know them
in the Milky Way may not be a representative sample of young star clusters
in galaxies in general. It has been known for many years that the 
Magellanic Clouds, and the LMC in particular, contain a number of young
``populous'' or ``massive'' clusters (hereafter YMCs) which have no 
counterpart in the Milky Way
(e.g.\ Hodge 1961). Conversely, the apparent \emph{paucity} of star clusters
in the irregular dwarf galaxy IC 1613 has presented an equally puzzling case
(van den Bergh 1979).  The presence of exceedingly luminous ``super star 
clusters'' (SSCs) in the nearby starburst galaxy M82 was first pointed out by 
van den Bergh (1971). Similar objects in NGC~1569 and NGC~1705 were noted
by Sandage (1978) and Arp \& Sandage (1985), and more recently 
HST observations have revealed large numbers of SSCs in many starburst 
galaxies, most notably in mergers like e.g.\ the ``Antennae'', NGC 4038/4039 
(Whitmore \& Schweizer 1995). Although the presence of SSCs/YMCs is often 
linked to starburst activity, it is less clear why they are
also present in some apparently normal galaxies (Kennicutt \& Chu 1988).  In 
this contribution I discuss the main results of a recent study of YMCs in a 
number of nearby late-type galaxies (mostly spirals), aiming at a better 
understanding of the differences between young cluster populations in 
different environments.

\section{Data}

  $UBVI$ and H$\alpha$ CCD imaging data for 22 nearby, mildly inclined 
spiral galaxies were collected with the Danish 1.54 m telescope at ESO / 
La Silla, the 2.56 m Nordic Optical Telescope at La Palma, Canary Islands 
and with the 3 m Shane reflector at Lick Observatory.  A YMC was defined as 
a point source without H$\alpha$ line emission, $B-V<0.45$, and with 
$M_V < -8.5$ for $U-B\ge-0.4$ and $M_V<-9.5$ for $U-B<-0.4$. These selection 
criteria ensure minimal contamination from Galactic foreground stars as well 
as individual luminous stars within the galaxies themselves. In some cases
we were also able to reidentify our cluster candidates on archive HST
images, confirming that most of the objects identified on the ground-based
images were indeed young clusters (Larsen 2000).  The adopted limit
in $B-V$ corresponds to an upper age limit of about 500 Myr, but the sample 
is biased towards younger clusters because of their generally higher 
luminosities.  In addition to our own ground-based observations, data for a 
number of other galaxies were compiled from the literature,
forming the basis 
for a comparison of the 
young cluster systems in a wide variety of environments.  
Details about the data reduction procedures, cluster selection criteria and 
a list of the galaxies are given in Larsen \& Richtler (1999; 2000).  

\section{Results}

  The 22 observed galaxies showed a wide variety in the properties
of their cluster systems, with the number of YMCs in each galaxy ranging from 
a handful or less (e.g.\ NGC 45, NGC 247, NGC 300) to more than 100 (M83, NGC 
6946). The left panel in Figure 1 shows the absolute $V$-band magnitude of the 
brightest cluster in each galaxy ($M_V^{\rm br}$) as a function of the host 
galaxy star formation rate (SFR), derived from IRAS far-infrared fluxes 
(Kennicutt 1998).  The $+$ markers indicate the 22 spiral galaxies, while 
the $*$ symbols denote literature data.  The plot shows a well-defined 
correlation between $M_V^{\rm br}$ and the host galaxy SFR. 
Interestingly, even IC~1613 (data from Wyder, Hodge and Cole 2000) fits 
quite nicely into the relation.  NGC~1569 and NGC~1705 represent possible
outliers, although it should be noted that the cluster ``systems'' in each 
of these two galaxies are dominated by only 1 or 2 very bright clusters 
(O'Connell, Gallagher, \& Hunter 1994).  

  From Figure 1, some galaxies evidently contain much more luminous
clusters than others, with $M_V^{\rm br}$ ranging between $-8.5$ and $-13$ 
for the spirals. For comparison, the brightest known open clusters in the 
Milky Way have $M_V\sim-10$ (e.g.\ $h$ and $\chi$ Persei, Schmidt-Kaler 1967) 
although a few even brighter clusters might hide in remote parts of the 
disk. Translating the $M_V$ magnitudes to cluster masses is 
non-trivial because the mass-to-light (M/L) ratios are very sensitive to 
age, differences in the stellar IMF etc. For a Salpeter IMF extending down 
to 0.1 M$_{\odot}$, the $V$-band luminosity per unit mass changes by nearly 
four magnitudes between an age of 5 Myr and 500 Myr (Bruzual \& Charlot 2001, 
in preparation), but for a 
typical cluster age of $\sim 20 \times 10^6$ years the $-8.5 < M_V < -13$ 
range corresponds to a mass range of very roughly 
$10^4$ M$_{\odot} <$ M $< 10^6$ M$_{\odot}$.

\begin{figure}
\begin{minipage}{65mm}
\epsfxsize=65mm
\epsfbox[82 373 545 715]{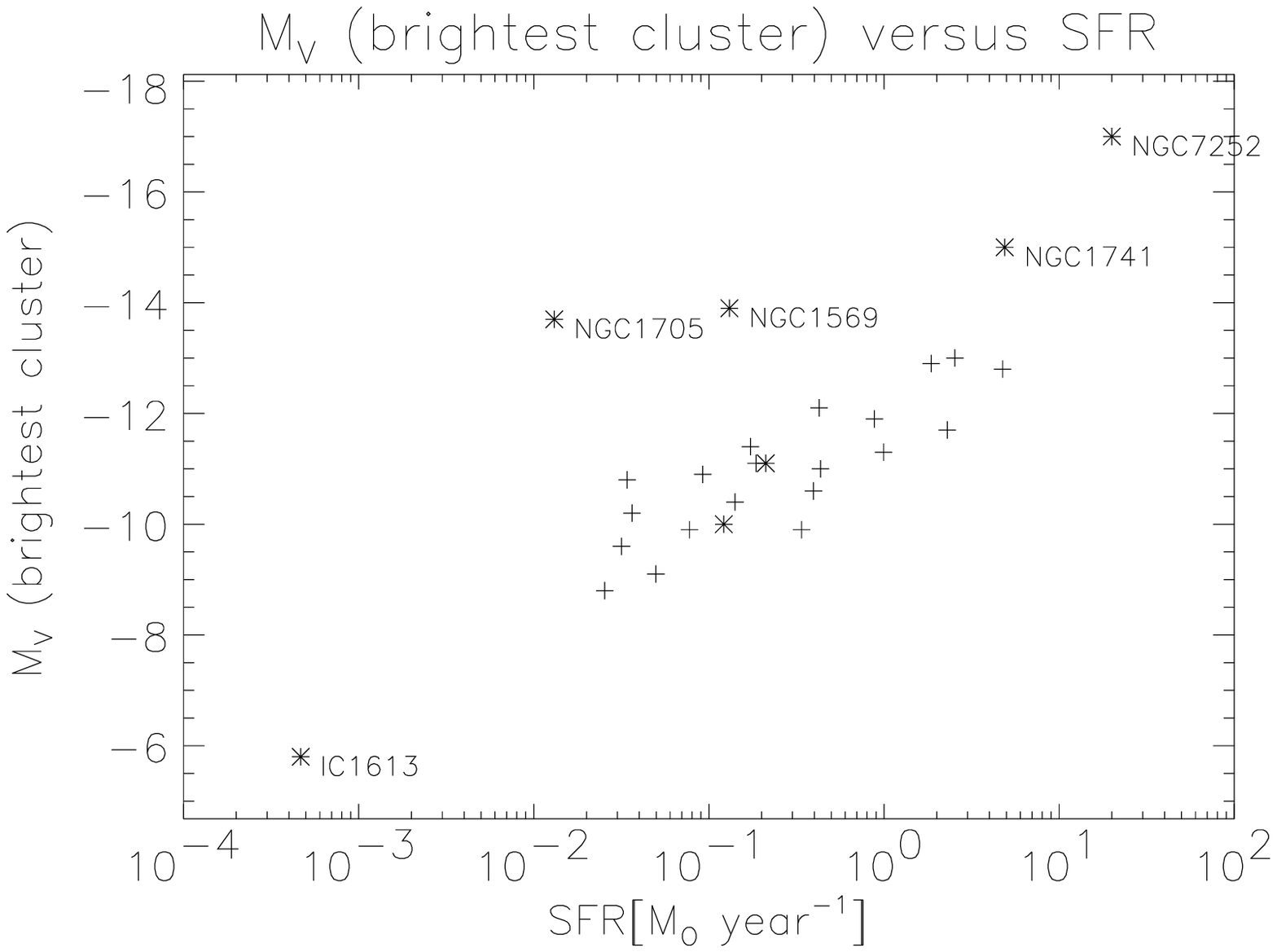}
\end{minipage}
\begin{minipage}{65mm}
\epsfxsize=65mm
\epsfbox[82 373 545 715]{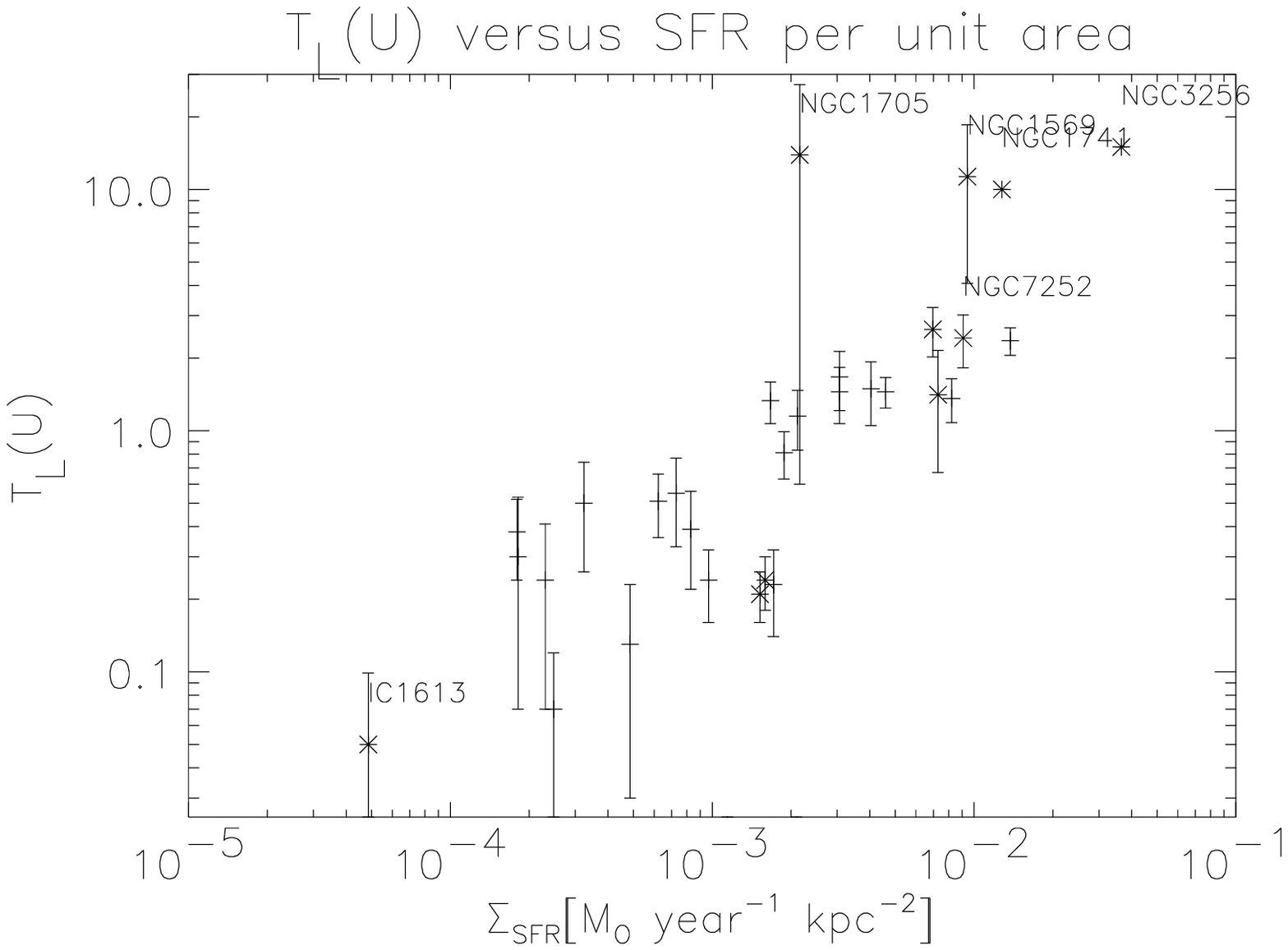}
\end{minipage}
\caption{Left panel: The magnitude of the brightest cluster, $M_V^{\rm br}$ as 
a function of host galaxy star formation rate. Right panel: Specific 
luminosity, $T_L(U)$, versus area-normalised star formation rate 
$\Sigma_{\rm SFR}$.}
\end{figure}

A relation of the type shown in the left-hand panel of Figure 1 might be 
expected just from sample statistics, since galaxies with high SFRs will 
also tend to have larger numbers of clusters and the likelihood of 
encountering very luminous clusters is higher in richer cluster systems,
unless the luminosity function is truncated.
  Another useful tool for studying young cluster populations, one that does 
not suffer from this effect, is the \emph{specific luminosity}, defined as 
\begin{equation}
  T_L = 100 \, \frac{L({\rm clusters})}{L({\rm galaxy})}
\end{equation}
  where $L$(clusters) is the total integrated luminosity of the cluster system 
and $L$(galaxy) is the luminosity of the host galaxy. Rather than using the
specific \emph{frequency}, as is customary in studies of old globular cluster 
populations, $T_L$ has the advantage of being less sensitive to incompleteness 
effects because it is dominated by the brightest clusters.  The right-hand 
panel of Figure 1 shows the $U$-band specific 
luminosity $T_L(U)$ (see Larsen \& Richtler 2000), now as a function of the 
area-normalised star formation rates of the host galaxies ($\Sigma_{\rm SFR}$).
Again, a quite well-defined correlation exists. As before, NGC~1569 and 
NGC~1705 have rather high $T_L(U)$ values for their star 
formation rates, but the large error bars that result from the poor statistics 
in these galaxies now make this deviation less striking. The general
impression from the figure is that $T_L(U)$ increases steadily as a function
of $\Sigma_{\rm SFR}$, with IC~1613 constituting one extreme endpoint of 
the relation, and active starbursts like NGC~1741 and NGC~3256 at the other.

\section{Discussion}

  First of all, one notes that it is very difficult to
make a meaningful division between galaxies with and without ``massive''
star clusters.  Both $M_V^{\rm br}$ and $T_L(U)$ show a steady progression 
with the host galaxy SFR and it appears that massive clusters (according 
to anyone's preferred definition) form \emph{whenever the host galaxy SFR 
is high enough}. Thus the very rich cluster systems in mergers and starburst
galaxies are naturally explained as due to the very high SFRs there, and it 
is not necessary to invoke special mechanisms which operate only in these 
environments to explain the presence of very luminous, young clusters,
other than those that triggered the starbursts in the first place.

Massive star clusters probably form a natural extension 
of the normal open cluster luminosity function (LF) to higher luminosities 
in galaxies with high SFRs, rather than being a distinct class of objects. 
In fact, van den Bergh \& Lafontaine (1984) have shown that extrapolation of 
the Milky Way open cluster LF to brighter magnitudes would yield a total 
of $\sim100$ objects with $M_V=-11$, which is clearly incompatible with the 
observations. They thus suggested a drop-off in the Milky Way open cluster 
LF somewhere in the range $-11 < M_V < -8$. If this drop-off occurs at
different magnitudes in different galaxies, this could strongly affect the
number of massive young clusters. Although our ground-based data did not
allow us to study the faint end of the LF for young clusters, some crude
estimates show that the observed number of clusters with 
$-12 < M_V < -10$ in YMC--rich galaxies like M51 are compatible with quite 
normal populations of ``open'' clusters and extrapolation of the open cluster 
LF to higher luminosities (Larsen 2000).

  The $U$-band specific luminosity is mostly sensitive
to the light from young stellar populations, and is therefore likely to
be an indicator of the relative fraction of stars forming in bound clusters,
relative to field stars. One then finds that this fraction \emph{increases} 
with the star formation rate, from less than 0.1\% in galaxies like IC~1613 
to more than 10\% in active starbursts.

  Eventually, understanding the formation of massive clusters will mean
understanding the physical processes in the interstellar medium and molecular 
clouds in which they are born. In the Milky Way, star clusters generally 
form within Giant Molecular Clouds (GMCs) with masses up to a few times 
$10^6$ M$_{\odot}$.  Although GMCs potentially have enough gas to form 
quite massive clusters, cluster formation in the Milky Way is evidently 
an inefficient process.  There are basically two ways to form higher-mass
clusters: Either GMCs in other galaxies are somehow able to 
convert a higher fraction of their mass into bound star clusters, or clusters
form with a constant efficiency in all galaxies and formation of YMCs
requires larger molecular clouds than the Milky Way GMCs.
The notion of such `Super-GMCs' (SGMCs) was conceived by Harris \& Pudritz
(1994), originally with the aim of explaining the formation of old
globular clusters in galactic halos, but we might now have a chance to test 
this idea by observing galaxies which are currently forming YMCs.  A few 
high-resolution studies of CO gas have been carried out for M51, M83 and the 
Antennae (Rand, Lord, \& Higdon 1999; Wilson et al.\ 2000), 
all of which are now known to contain rich YMC populations.  These studies 
have indeed detected ``Giant Molecular Associations'' 
with masses of $10^7 - 10^8$ M$_{\odot}$, or 1 -- 2 
orders of magnitude higher than for Milky Way GMCs.  Whether or not SGMCs 
are the birthsites of YMCs remains to be verified, and it is possible that 
these complexes might resolve into smaller subunits when examined at higher 
resolutions. In any case, more high-resolution studies of molecular gas in 
galaxies with YMC populations are likely to provide important insight into 
their formation.

\section{Conclusions}

\begin{itemize}
  \item ``Massive'' clusters can form in many different environments.
        Although mergers represent one efficient way of providing the 
	required high star formation rates, they are evidently not the only 
	way to form massive clusters and processes that operate only
	in mergers or otherwise disturbed galaxies (like large-scale 
	cloud-cloud collisions) are unlikely to be the primary mechanism 
	responsible for YMC formation.  It seems more likely that all star 
	clusters (very likely even globular clusters in galactic halos) form 
	by the same basic physical mechanism. 

  \item The distinction between ``open'', ``massive'' and ``super'' 
        clusters may turn out to be largely an artificial one. If we
        had been living in a different galaxy, chances are our classification
        of stellar clusters would have been different, too.

  \item 
        $M_V^{\rm br}$ and 
	$T_L(U)$ both increase gradually as a function
        of the host galaxy SFR. At one extreme of the
        relation are galaxies like IC~1613 with very low SFRs and 
        correspondingly feeble cluster systems. At the other extreme are
        starbursts and merger galaxies with their very high SFRs and large 
        numbers of highly luminous clusters. ``Normal'' galaxies fall in 
        between these extremes, but still show significant variations in 
	$M_V^{\rm br}$ and $T_L(U)$.  Some nearby spirals like M51, M83 and 
	NGC~6946 contain young clusters that are almost as luminous as those 
	in the ``Antennae'' galaxies.
\end{itemize}

\acknowledgments
This work was partly supported by NSF grant AST9900732 and by a travel
grant from the American Astromical Socitety.

\end{document}